An easy method for calculating the motion of celestial bodies perturbed in any manner avoiding astronomical computations.

*E348 – Methodus facilis motus corporum coelestium utcunque perturbatos ad rationem calculi astronomici revocandi.*

Translated and Annotated[†]
by

Sylvio R. Bistafa[*]
June 2019

Foreword

As earlier as the 1730's and until his death in 1783, Euler wrote several papers on celestial mechanics, which are generally characterized by rather lengthy and intricate astronomical computations. A typical example of the works of this period is E187 – 'Theory of the motion of the moon which exhibits all its irregularities', which was originally published as a book in 1753, and contains 18 chapters written in Latin. As revealed by its title itself, E348 of 1768 was written with the goal of alleviating the astronomical computations in a typical celestial three-body problem represented by Sun, Earth and Moon. In this work, Euler's approach consists of two parts: geometrical and mechanical. The geometrical part contains most of the analytical developments, in which Euler makes use of Cartesian and spherical trigonometry as well — the latter not always in a clear enough way. With few sketches to show the geometrical constructions envisaged by Euler — represented by several geometrical variables —, it is a hard to follow publication. The Translator, on trying to clear the way to the non-specialized reader, used the best of his abilities to add his own figures to the translation. In the latter part of the work, Euler particularizes his developments to the Moon, ending up with eight coupled differential equations for resolving the perturbed motion of this celestial body, which makes his claim of an "easy method" as being rather fallacious. Despite showing great analytical skills, Euler did not give indications on how this system of equations could be solved, which renders his efforts practically useless in the determination of the variations of the nodal line and inclination of the Moon's orbit.

Author
L. Euler

I.

---

[†] The translator used the best of his abilities and knowledge to make this translation technically and grammatically as sound as possible. Nonetheless, interested readers are encouraged to make suggestions for corrections as they see fitting.
[*] Corresponding address: sbistafa@usp.br

Although I have often attacked the investigation on how the motions of celestial bodies are perturbed due to their mutual action, most of the time, I have incurred in rather lengthy and laborious calculations, which, however, after many digressions, could be reduced to simpler formulas. However, the cause of this prolixity is due to a multiplicity of elements, which are necessary to introduce into the calculation, that is: not only for all the determinations, which are related to the motion of the perturbing bodies, should be examined, but also the perturbation of its own motion, in so far as if it is not in the same plane, it demands various elements, to which is a custom among Astronomers to consider variations originated in the nodal line[1] and inclination of the orbit. But in case all these considerations are simultaneously included into the calculation, it is really not worthwhile, because they will give rise to much trouble and confusion, to which no other remedy is seen to exist, unless all the elements are carefully distinguished, and all the operations are established so that no more elements are admitted into them, than those that are necessary to consider.

II.

The principal part of this research is related to mechanics, when the perturbation of the motion by the forces of the disturbing body should be defined; thus mechanical principles are provided, and from them the location of the body, whose motion is seeked, can be conveniently determined at any time by three mutually orthogonal coordinates; truly the other part scarcely requires a swifter development, with which the location is firstly determined, and should be reduced to the acceptable practice in Astronomy, in which it is common to naturally express the different locations in the sky by longitude and latitude. And also in this second part, in which will be allowed to recall the geometry, being correct to distinguish a priori that all mechanics is due to it, and I should observe that these two parts can furthermore not only be conveniently separated, but also to be able to deal with both matters in a much easier way than if we wished to deal jointly with both of them. However, it is seen that the geometrical mechanical investigation should precede, nonetheless it is possible to begin with the geometrical part in a neat way, with none impediment, from the location of the body, which motion we seek to obtain, as it were known, and that we identify by three coordinates. This inversion of the methodology is thus seen to follow, so that the development of the geometrical part gives much important support, with which the work in the calculation of the mechanical part that follows next will be considerably alleviated. We follow this method certainly with great advantage, provided that we handle the geometrical part without introducing into the calculation quantities related to the disturbing forces.

GEOMETRICAL PART

III.

Therefore, I assume that the motion of the body $Z$ is to be determine and that, as usual, readily defined by mechanical principles. Certainly, firstly the motion in relation to certain point $A$, which is considered fixed, even if it happened that the said point is used as reference to the circular motion, then, next, a certain plane is considered traversing through this point and equally fixed, which is represented by the plane of the figure itself (Fig. 1), in which it is drawn a fixed line $AB$, and at any time, the location of the body $Z$ is thus defined by the three mutually orthogonal coordinates $AX$, $XY$ and $YZ$, so that first, from the $Z$ location, the perpendicular $ZY$ to that plane is drawn, further on, truly from $Y$, the normal $YX$ is guided in the direction of the line $AB$. Then, let us call the following three coordinates:

$$AX = X, XY = Y \text{ and } YZ = Z$$

---

[1] The nodal line is a line that joins the ascending node and the descending node of an orbit. It marks the intersection of the orbital plane and some reference plane, usually the ecliptic.

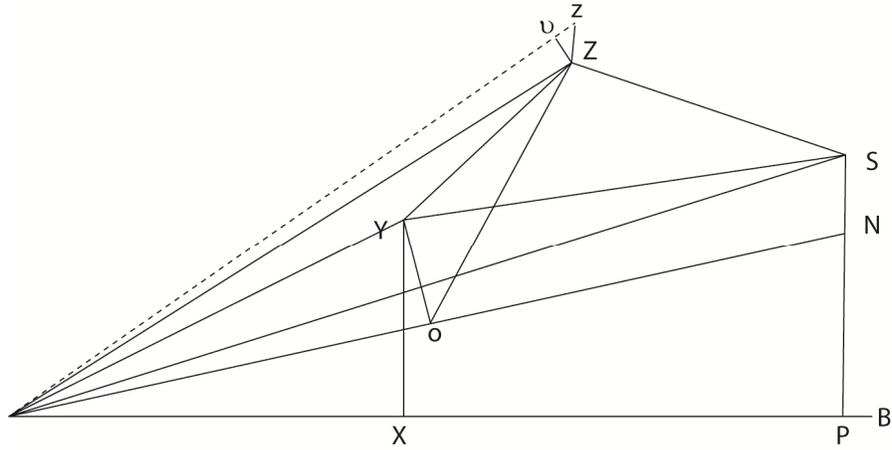

Figure 1

which values at any elapsed time $= t$ are considered to be known. Then, consequently, the distance of the body $Z$ to the fixed point $A$ is immediately obtained, which, for brevity, it is indicated by $AZ = v$, and then, $v^2 = X^2 + Y^2 + Z^2$.

Then, we conceive that during the time $dt$ the body advances from $Z$ to $z$, such that $Az = v + dv$, and the elementary angle $ZAz = d\emptyset$, which, in the mean time, the body $Z$ is seen to complete in its orbit around $A$, resulting in $Zz = \sqrt{dv^2 + v^2 d\emptyset^2}$, whereas, according to the coordinate elements we have that $Zz = \sqrt{dX^2 + dY^2 + dZ^2}$, whence

$$dv^2 + v^2 d\emptyset^2 = dX^2 + dY^2 + dZ^2$$

in this way, this exposes how the elementary angle $d\emptyset$ can be conveniently expressed by the coordinates, which will be soon succinctly shown.

IV.

A certain plane is defined by the segment $Zz$ and point $A$, in which the body $Z$ is, in fact, considered to move: this plane will cut somewhere the fixed plane of the figure; then this intersection is constructed along the line $AN$, which is called the nodal line in Astronomy, and which variation, due to the motion perturbation of the body $Z$, should be investigated above all: next, it is also convenient to note, that the angle which the plane $NAZ$ is inclined towards the fixed plane, which in Astronomy is simply called inclination, and because of the perturbation of the motion, can undergo remarkable alterations. Then, next, let us consider these new elements:

Longitude of the nodal line or angle $BAN = \psi$

Inclination of the orbit to the fixed plane $= \omega$

and argument of latitude or angle $NAZ = \sigma$

which we rename according to the coordinates, so from $Y$, as well as from $Z$, let us draw the normals $YO$ and $ZO$ to the nodal line $AN$, such that the angle $YOZ$ will have the same inclination of $\omega$.

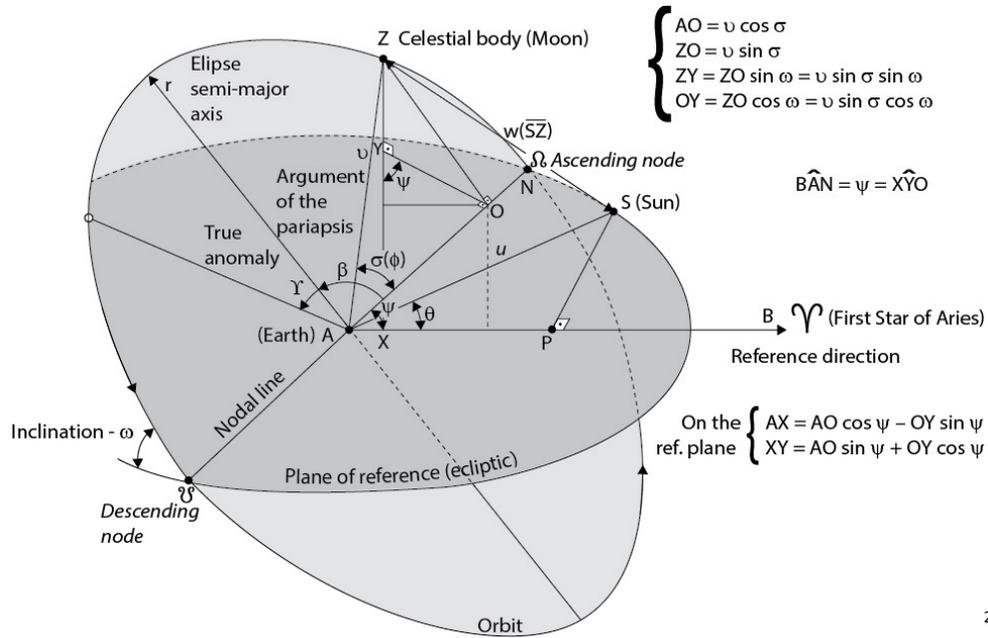

$$\begin{cases} AO = \upsilon \cos \sigma \\ ZO = \upsilon \sin \sigma \\ ZY = ZO \sin \omega = \upsilon \sin \sigma \sin \omega \\ OY = ZO \cos \omega = \upsilon \sin \sigma \cos \omega \end{cases}$$

$$\widehat{BAN} = \psi = \widehat{XYO}$$

On the $\begin{cases} AX = AO \cos \psi - OY \sin \psi \\ XY = AO \sin \psi + OY \cos \psi \end{cases}$
ref. plane

Then, considering that the angle $NAZ = \sigma$, and that the distance $AZ = v$, we will have that:

$$AO = v \ cos\sigma \quad \text{and} \quad ZO = v \ sin\sigma$$

and further

$$ZY = v \ sin\sigma \ sin\omega \quad \text{and} \quad OY = v \ sin\sigma \ cos\omega$$

and since the angle $BAN = \psi = XYO$, we conclude that:

$$AX = v \ cos\sigma \ \cos \psi - v \ sin\sigma \ cos\omega \ sin \ \psi$$

and $AY = v \ cos\sigma \ sin\psi + v \ sin\sigma \ cos\omega \ cos \ \psi$.

Therefore, our three coordinates are thus defined:

$$X = v \ (cos\sigma \ \cos \psi - \ sin\sigma \ cos\omega \ sin \ \psi \ )$$

$$Y = v \ (cos\sigma \ sin \ \psi + \ sin\sigma \ cos\omega \ cos \ \psi \ )$$

and $Z = v \ sin\sigma \ sin\omega$,

where it should be noted that the tangent of the angle $NAY = tan\sigma \ cos\omega$, which is an angle called the longitude of the point $Z$ to the node.

<div align="center">V.</div>

Since in Astronomy the angle $BAY$ reveals the longitude, truly the angle $ZAY$ is the latitude of point $Z$, which relates to the plane of the ecliptic, and the line $AB$ extended to the First Star of Aries; each one of these denominations can be used in a wider sense: we then have

<div align="center">longitude of the point $Z$ or angle $BAY = \psi + NAY$</div>

<div align="center">considering that the tangent of the angle $NAY = tan\sigma \ cos\omega$</div>

<div align="center">for the latitude or truly for the angle $ZAY$ we will have</div>

---

$$sin\,ZAY = \frac{ZY}{AZ} = sin\sigma\,sin\omega$$

where the same formulas are usually obtained from the spherical trigonometry. Certainly, in the spherical surface with center in $A$, the maximum circle $BNY$ represents a fixed plane, and the point $B$ at infinite, and from which the longitude is calculated (Fig. 2). Furthermore, be $N$ the node and $NZ$ the orbit to which now the motion of the body $Z$ is referred to, then, from $Z$, in the direction of the circle $BNY$, the normal arch $ZY$ is drawn; once this is done, the arch $BNY$ shows the longitude, and truly, the arch $ZY$ the latitude of the point $Z$, and in relation to these we have that:

arch $BN$, or angle of the node $= \psi$

angle $ZNY$ or inclination $= \omega$

and arch $NZ$ or argument of the latitude $= \sigma$

from these, the solution of the right spherical triangle $NYZ$ gives

$$sin\,ZY = sin\sigma\,sin\omega \qquad and \qquad tan\,NY = tan\sigma\,cos\omega,$$

exactly as before.

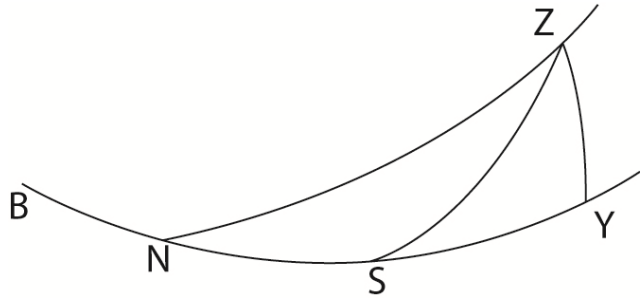

Figure 2

VI.

However, for the nodal line and the inclination be both variables; since both the point $Z$ and $z$ belong to the same plane $NAZ$, then by differentiation, the point $Z$ should come to $z$, and because the angle $BAN = \psi$ and the inclination $\omega$ are considered constants, of course as long as the angle $NAZ = \sigma$, the elementary angle $ZAz = d\emptyset$ is assumed to increase, such that $d\sigma = d\emptyset$. However, to came to the same point $z$ in another way, it is necessary that the nodal line and the inclination be both considered variables, since the point $z$ should also have a tendency to a diversified orbit, and then, the differential $d\sigma$ should not be considered to be equal to $d\emptyset$ itself, but it should attain its proper value, which at the same time depends on the variation of the orbit. Therefore, since this double differentiation should lead to the same equations that we will obtain next, for which certain relations between the variations originated in the orbit will be defined, which will provide a maximum usage in a subsequent calculation. In fact, it possesses not only the differentiations of the local coordinates themselves, but also of the quantities thence derived, such as:

$$\frac{X}{Z} = \frac{cos\sigma\,cos\,\psi}{sin\sigma\,sin\omega} - \frac{cos\sigma\,sin\,\psi}{sin\omega} \quad and \quad \frac{Y}{Z} = \frac{cos\sigma\,sin\,\psi}{sin\sigma\,sin\omega} + \frac{cos\omega\,cos\,\psi}{sin\omega}$$

which ought, therefore, to provide the same results obtained in both ways for the double differentials.

VII.

Therefore, firstly assuming that the angles $\psi$ and $\omega$ are constants and that $d\sigma = d\emptyset$, the differentials are:

$$d\left(\frac{X}{Z}\right) = \frac{-d\emptyset \cos\psi}{\sin^2\sigma \ \sin\omega} \qquad \text{and} \qquad d\left(\frac{Y}{Z}\right) = \frac{-d\emptyset \ \sin\psi}{\sin^2\sigma \ \sin\omega}.$$

For the other differentiation [considering that $\psi$ and $\omega$ are also variables], it should be firstly noted that:

$$\frac{X}{Z}\cos\psi + \frac{Y}{Z}\sin\psi = \frac{\cos\sigma}{\sin\sigma \ \sin\omega} \qquad \text{and} \qquad \frac{Y}{Z}\cos\psi - \frac{X}{Z}\sin\psi = \frac{\cos\omega}{\sin\omega}$$

which, in the usual way of differentiating gives:

$$\cos\psi \ d\left(\frac{X}{Z}\right) + \sin\psi \ d\left(\frac{Y}{Z}\right) + d\psi\left(\frac{Y}{Z}\cos\psi - \frac{X}{Z}\sin\psi\right) = \frac{-d\sigma}{\sin^2\sigma \ \sin\omega} - \frac{d\omega \ \cos\sigma \ \cos\omega}{\sin\sigma \ \sin^2\omega}$$

$$\cos\psi \ d\left(\frac{Y}{Z}\right) - \sin\psi \ d\left(\frac{X}{Z}\right) - d\psi\left(\frac{X}{Z}\cos\psi + \frac{Y}{Z}\sin\psi\right) = \frac{-d\omega}{\sin^2\omega},$$

whence, when the former values are substituted results in[3]:

$$\frac{-d\emptyset \cos^2\psi}{\sin^2\sigma \ \sin\omega} - \frac{d\emptyset \ \sin^2\psi}{\sin^2\sigma \ \sin\omega} + \frac{d\psi \ \cos\omega}{\sin\omega} = \frac{-d\sigma}{\sin^2\sigma \ \sin\omega} - \frac{d\omega \ \cos\sigma \ \cos\omega}{\sin\sigma \ \sin^2\omega}$$

$$-\frac{d\emptyset \ \sin\psi \ \cos\psi}{\sin^2\sigma \ \sin\omega} + \frac{d\emptyset \ \sin\psi \ \cos\psi}{\sin^2\sigma \ \sin\omega} - \frac{d\psi \ \cos\sigma}{\sin\sigma \ \sin\omega} = \frac{-d\omega}{\sin^2\omega},$$

which are reduced into these:

$$-\frac{d\emptyset}{\sin^2\sigma \ \sin\omega} + \frac{d\psi \ \cos\omega}{\sin\omega} = \frac{-d\sigma}{\sin^2\sigma \ \sin\omega} - \frac{d\omega \ \cos\sigma \ \cos\omega}{\sin\sigma \ \sin^2\omega}$$

$$-\frac{d\psi \ \cos\sigma}{\sin\sigma \ \sin\omega} = \frac{-d\omega}{\sin^2\omega} \quad \text{or} \quad \frac{d\omega}{\sin\omega} = \frac{d\psi}{\tan\sigma},$$

which when substituted into the previous expression gives

$$\frac{d\sigma - d\emptyset}{\sin^2\sigma \ \sin\omega} = \frac{-d\psi \ \cos\omega}{\sin\omega} - \frac{d\psi \ \cos^2\sigma \ \cos\omega}{\sin^2\sigma \ \sin\omega} = \frac{-d\psi \ \cos\omega}{\sin^2\sigma \ \sin\omega}$$

or $d\psi \ \cos\omega = d\emptyset - d\sigma$.

VIII.

Whence, therefore, we first learned that the variation in the inclination of the orbit giving rise to $d\omega$ is such that it always depends on the variation of the nodal line $d\psi$ that is, $d\omega = \frac{d\psi \ \sin\omega}{\tan\sigma}$; or the increment in the inclination will be due to the promotion of the nodal line, as the sine of the inclination to the tangent of the argument of the latitude; whence the following consequences can be drawn from it:

1º. If the argument of the latitude $\sigma$ is zero or 6s where the latitude is zero, in the mean time the nodal line will tend to remain at rest the more the inclination is varied[4].

2º. If the argument of the latitude $\sigma$ is 3s or 9s where the latitude is zero, or $\tan\sigma = \infty$ where the latitude is maximum, then the inclination will not vary; regardless if in the mean time the nodal line progresses or regresses.

---

[3] No justification is given by Euler for this equality.
[4] The ecliptic was in the past divided into 12 signs, each subdivided into 30 degrees.

3º. If the argument of the latitude $\sigma$ is contained between the limits 0s and 3s or between 6s and 9s, that is, while the latitude increases, then the inclination $\omega$ increases, if indeed the nodal line advances, but if it retreats, the inclination diminishes.

4º. If the argument of the latitude $\sigma$ is contained between the limits 3s and 6s or between 9s and 12s, that is, while the latitude decreases, then the advancement of the nodal line inclination lessens, and in reality it ceases to be increased.

<div align="center">IX.</div>

Next, it should be observed that the increase of the argument of the latitude $\sigma$ promoted in its own orbit is not equal to the element $d\emptyset$, unless the nodal line stays immovable; since we found that $d\sigma = d\emptyset - d\psi \, cos\omega$, with the exception in the case when the inclination $\omega$ were a right angle. These phenomena will be expressed more clearly by using the spherical trigonometry (Fig. 3). If in fact, as before, the circle $BNY$ represents a fixed plane, which the motion of the point $Z$ is referred to, so that its present motion takes place according to the circle $NZ$, such that $BN = \psi$, $YNZ = \omega$ and the arch $NZ = \sigma$, whereas after the point $Z$ has progressed through the element $Zz = d\emptyset$, and with its motion taking place according to the circle $nz$, the promotion of the nodal line will be given by $Nn = d\psi$, the inclination is transformed into $Ynz = \omega + d\omega$, and the argument of the latitude into $nz = \sigma + d\sigma$. Accordingly, the elemental arch $nv$ is drawn normal to $NZ$, and then we will have that $Nv = d\psi \, cos\omega$ and $nv = d\psi \, sin\omega$; thence it is deduced that $Zn = \sigma - d\psi \, cos\omega$, and on this account $nz = \sigma - d\psi \, cos\omega + d\emptyset = \sigma + d\sigma$ and consequently, $d\sigma = d\emptyset - d\psi \, cos\omega$, as before; however, at the same time we see that because $d\psi \, cos\omega = d\emptyset - d\psi = Nv$, the expression $d\emptyset - d\psi$ exhibits the promotion of the nodal line in the orbit itself, because since the node was in the point $N$ of the orbit $NZ$; it has now been transferred to the point $n$ or to $v$. In addition, from the spherical triangle $NnZ$, we have that:

$$sin\omega : sin(\omega + d\omega) = sin(\sigma - d\psi \, cos\omega) : \, sin\sigma \ ^{5}$$

or

$$sin\omega : sin\omega + d\omega \, cos\omega = sin\sigma - d\psi \, cos\sigma \, cos\omega : sin\sigma$$

and dividing by

$$sin\omega : d\omega \, cos\omega = sin\sigma - d\psi \, cos\sigma \, cos\omega : d\psi \, cos\sigma \, cos\omega \ ^{6}$$

gives

$$d\omega \, sin\sigma = d\psi \, sin\omega \, cos\sigma \ ^{7} \qquad \text{or} \qquad \frac{d\omega}{sin\omega} = \frac{d\psi}{tan \, \sigma}$$

exactly as before[8].

---

[5] The application of the Law of Sines to the spherical triangle $NnZ$ does not give this expression.
[6] It is not known where this expression comes from.
[7] The claimed division does not lead to this expression.
[8] From the above observations, it appears that this result was forced by Euler. Nonetheless, this has no further consequences, since the same expression was obtained before by another method in § 7.

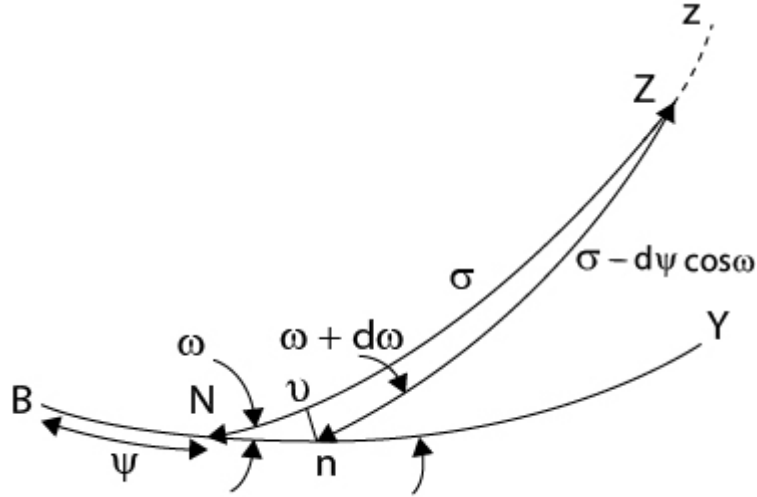

Figure 3

X.

Moreover, if the differential formulas found in § 7. are unfolded they give

$$\frac{XdZ - ZdX}{Z^2} = \frac{d\emptyset \, cos\psi}{sin^2\sigma \, sin\omega} \quad \text{and} \quad \frac{YdZ - ZdY}{Z^2} = \frac{d\emptyset \, sin\psi}{sin^2\sigma \, sin\omega}$$

and since $Z = v \, sin\sigma \, sin\omega$, we find these very fitting formulas:

$$XdZ - ZdX = v^2 d\emptyset \, sin\omega \, cos\psi \quad \text{and}$$

$$YdZ - ZdY = v^2 d\emptyset \, sin\omega \, sin\psi.$$

Next, we eliminate $dZ$, firstly multiplying by $Y$, and then by $X$, and once the product is taken, it will give

$$Z(XdY - YdX) = v^2 d\emptyset \, sin\omega(Ycos\psi - Xsin\psi).$$

However, as we saw in § 7, $Ycos \, \psi - Xsin \, \psi = \frac{Z \, cos\omega}{sin\omega}$, this formula is reduced to a much simple expression

$$XdY - YdX = v^2 d\emptyset \, cos\omega.$$

These formulas when combined with those then found at the beginning, namely

$$X^2 + Y^2 + Z^2 = v^2 \quad \text{and} \quad dX^2 + dY^2 + dZ^2 = dv^2 + v^2 d\emptyset^2$$

will be used with maximum advantage in the mechanical part, to the coordinates derived from the calculations, such that it provides to the next quantities of this sort, which use are retaken in Astronomy.

XI.

However, those reductions of the Geometry clearly show that it makes no difference to which point $A$ and fixed plane $BAY$ (see Fig. 1) we wish to refer the motion of the point $Z$. In fact, if we envisage an astronomical use, it is of most interest, not only in which way the point $A$, as the center of motion is taken into account, but also that plane, to which the motion of the point $Z$ is referred to by longitude and latitude, because from this will depend the chief simplicity of the determination. To this end, it is necessary to consider how that choice should be made, together with the artifices, which so far have been devised, and only then, it can be used with some success, since the motion of the body, which is sought, should not disagree very much from the laws of Kepler, on account that the perturbations hand been very small. Moreover, when the motion is thus compared, such that the areas described around any point are nearly proportional to time, then this point is most suitable to be considered as that fixed point $A$. When that happens, if among the forces driving the body, one far exceeds the remaining, to that point this force should be directed to, then point $A$ will be suited to be accepted: therefore, if the question would be related to perturbations of a certain chief planet or of a comet, then point $A$ will be most suitable taken in the center of the Sun: if however, the perturbations in the motion of the Moon, or those made in another secondary planet should be defined; then it is proper be considered point $A$ in the center of the Earth or [in the center] of the primary planet, such that the force of the body declared in $Z$ impelling it to $A$, much exceeds the remaining forces to which this body is simultaneously driven.

XII.

So, If the body $Z$ has been solicited by just one principal force, the body will be revolved regularly around point $A$ in a conical section, perpetually in the same plane, such that no matter in what way the fixed plane $BAY$ is chosen, neither on how the nodal line nor the inclination and any mutation has been ever originated; however, meanwhile, the calculation, without doubt, has turned out very simple, if the fixed plane is chosen in the same plane of the motion. Truly, if the motion is disturbed by another celestial body, which motion is indeed also necessary to be assumed known in this investigation, the fixed plane can most conveniently be assumed as being congruent with the orbit of that disturbing body. Thus, if the perturbations of the Moon originated by the Sun are sought, the ecliptic plane, in which the Sun is seen to move from the Earth, as it [the Earth] were the center of the motion $A$, will render the fixed plane $BAY$, and no matter how the perturbation from another body is brought about, this plane, in which this body is seen to move from the center of motion $A$, should be selected. Yet, if this body itself is not moved in the same plane, then some medium plane can be most conveniently adopted; but the effort to adapt the calculation to this case is hardly considered, but, if its employment will become indeed necessary, it can be easily provided.

MECHANICAL PART

XIII.

For the handling of the mechanical part, three bodies should be considered (see Fig. 1). The first, is the one which is putted in the center of the motion $A$, which exerts the main force in the body $Z$, which motion we investigate, such as it appears to an observer located in the very point $A$, let us then call the mass of the body positioned in this point $= A$.

The other body, by which action the motion of the body $Z$ is perturbed, that we assume is moving in any manner in the fixed plane $BAY$ itself, such that its location can be assigned at any time. Be the mass of this body $= B$, and that now, in fact, it dwells in $S$, such that its distance to the central body $AS = u$, and the longitude or angle $BAS = \theta$, whence, from $S$, the perpendicular $SP$ is drawn to the fixed line $AB$, and be $AP = u \, cos\theta$ and $PS = u \, sin\theta$.

The third body is the one in $Z$ itself, which motion we are looking for, be its mass $= C$, and as before, we put its distance to the center of motion $AZ = v$, and calling the three orthogonal coordinates $AX = X$, $XY = Y$ and $YZ = Z$, which we obtain from the calculation by introducing the following elements:

Finally, we consider that during the infinitesimal time $dt$, the elementary angle $ZAz = d\emptyset$ is completed by the body $Z$. On the other hand, the relationships of these elements will be taken from the geometrical part.

XIV.

Since the body $Z$ is driven to $A$ by a force $= \frac{A}{v^2}$, certainly, $A$ is attracted to $Z$ by a force $= \frac{C}{v^2}$, so that for the point $A$ could be considered at rest, the body $Z$ should be declared to be attracted to $A$ by a force $= \frac{A+C}{v^2}$,[10] which once resolved according to the directions of the three coordinates give the following forces: according to $XA := \frac{A+C}{v^3} \cdot X$; according to $YX := \frac{A+C}{v^3} \cdot Y$; according to $ZY := \frac{A+C}{v^3} \cdot Z$. Thereafter, for the force with which the body $Z$ is attracted towards $S$, we have: firstly, let us call, for simplicity, the distance $SZ = w$, such that the force $ZS$ is $= \frac{B}{w^2}$, which can be readily decomposed into the forces: according to $ZY := \frac{B}{w^3} Z$, and according to $YS := \frac{B}{w^3} YS$, and from this, since $XP = u \cos\theta - X$ and $PS - XY = u \sin\theta - Y$, we have the forces: according to $XP := \frac{B}{w^3}(u \cos\theta - X)$ and according to $XY := \frac{B}{w^3}(u \sin\theta - Y)$.[11]

Finally, because body $A$ is driven to $S$ by a force $= \frac{B}{u^2}$, the following components will be contrarily translated to $Z$: according to $XA := \frac{B}{u^2} \cos\theta$, and according to $YX := \frac{B}{u^2} \sin\theta$, which once collected with the other forces acting on the body $Z$ give:

1º. Force according to $XA := \frac{A+C}{v^3} X + \frac{B}{w^3}(X - u \cos\theta) + \frac{B}{u^2} \cos\theta$

2º. Force according to $XY := \frac{A+C}{v^3} Y + \frac{B}{w^3}(Y - u \sin\theta) + \frac{B}{u^2} \sin\theta$

3º. Force according to $ZY := \frac{A+C}{v^3} Z + \frac{B}{w^3} Z$,

which should be proportional to the acceleration of the body $Z$ in the same directions, and considering a constant element of time we have:

---

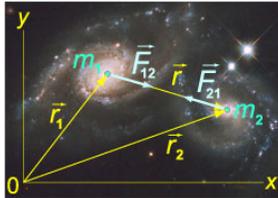

In a system of two bodies, the attraction force $F_{12}$ of the second body acts on the first body of mass $m_1$. Similarly, the attraction force $F_{21}$ of the first body acts on the second body of mass $m_2$. Both forces $F_{12}$ and $F_{21}$ are equal and directed along $r$, where $r = r_2 - r_1$. From Newton's second law, we can write the following differential equations describing the motion of each body: $m_2 \frac{d_{r_2}^2}{dt^2} = -G \frac{m_1 m_2}{r^3} r$ or $\frac{d_{r_1}^2}{dt^2} = G \frac{m_2}{r^3} r$, $\frac{d_{r_2}^2}{dt^2} = -G \frac{m_1}{r^3} r$, where $G$ is the gravitational constant. It follows from the last two equations that $\frac{d_{r_1}^2}{dt^2} - \frac{d_{r_2}^2}{dt^2} = G \frac{m_2}{r^3} r + G \frac{m_1}{r^3} r$, and then, $\frac{d_r^2}{dt^2} = -G \frac{m_1 + m_2}{r^3} r$. For $m_1 = A$, $m_2 = C$, $r = v$, and when the two bodies are collinear with $O$, results in $\frac{d_v^2}{dt^2} = -G \frac{A+C}{v^2}$, where the minus sign means that these forces tend to shorten the distance between the two bodies.

[11] These are the components of the force according to $YS$, projected in the directions $AX$ and $XY$.

$$ddX = -\propto dt^2 \left[\frac{A+C}{v^3}X + \frac{B}{w^3}X - Bu\,cos\theta\left(\frac{1}{w^3} - \frac{1}{u^3}\right)\right]$$

$$ddY = -\propto dt^2 \left[\frac{A+C}{v^3}Y + \frac{B}{w^3}Y - Bu\,sin\theta\left(\frac{1}{w^3} - \frac{1}{u^3}\right)\right]$$

$$ddZ = -\propto dt^2 \left[\frac{A+C}{v^3}Z + \frac{B}{w^3}Z\right],$$

where the constant $\propto$ depends on each particular type of motion, and can be defined from the apparent motion of the Sun.

<div align="center">XV.</div>

However, before we consider these formulas further, we should precisely define the distance $SZ = w$, to introduce it afresh into the calculation. Since we have that:

$$SZ^2 = ZY^2 + XP^2 + (PS - XY)^2 \quad {}^{12}$$

then,

$$w^2 = Z^2 + X^2 + Y^2 + u^2 - 2uX\,cos\theta - 2uY\,sin\theta,$$

which because $X^2 + Y^2 + Z^2 = v^2$, can be reduced to:

$$w^2 = v^2 + u^2 - 2u(X\,cos\theta + Y\,sin\theta).$$

Moreover, introducing the expressions for $X$ an $Y$ found in §4. above:

$$X\,cos\theta + Y\,sin\theta = v[cos\sigma\,cos(\theta - \psi) + sin\sigma\,cos\omega\,sin(\theta - \psi)],$$

where the angle $\theta - \psi$ expresses the distance of the disturbing body $S$ to the nodal line or angle $NAS = \theta - \psi$, so that

$$w^2 = v^2 + u^2 - 2vu[cos\sigma\,cos(\theta - \psi) + sin\sigma\,cos\omega\,sin(\theta - \psi)].$$

In fact, if now for brevity we call the angle $SAZ = \mu$, which denotes the distance of the body $Z$ to the disturbing body $S$ as seen from $A$, because $AZ = v$ and $AS = u$, it is also true that

$$w^2 = v^2 + u^2 - 2vu\,cos\mu,$$

whence, it will be concluded that:

$$cos\sigma\,cos(\theta - \psi) + sin\sigma\,cos\omega\,sin(\theta - \psi) = cos\mu$$

which can be easily proved by spherical trigonometry[13]. As seen in Fig. 2, since because $BN = \psi$, $NZ = \sigma$, and the angle $YNZ = \omega$, if $BS = \theta$, then $NS = \theta - \psi$, and in the spherical triangle the side $SZ = \mu$ is determined in this very way from the sides $NZ = \sigma$, $NS = \theta - \psi$ with the intercepted angle $ZNS = \omega$.

---

[12] The first term in the second hand-side of this expression was incorrectly written as $XZ^2$ in the original manuscript.
[13] This result comes from the application of the law of cosines to the spherical triangle of the figure.

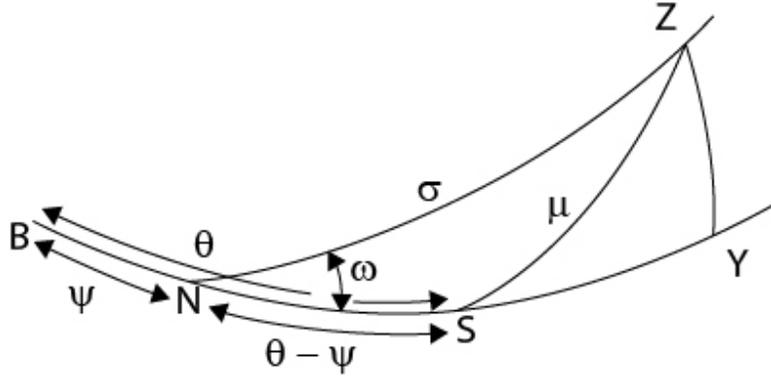

XVI.

Although the three equations deducted from the principles of mechanics are enough for all determinations, because all the craft in them is certain, just as it enable us to derive them in a very suitable manner. Nonetheless, it is first required to offer at once the calculation of the multiplications: the first [equation] by $2dX$, the second [equation] by $2dY$ and the third [equation] by $2dZ$, and in only one [equation] they ought be gathered; since as we saw above

$$dX^2 + dY^2 + dZ^2 = dv^2 + v^2 d\emptyset^2$$

$$\text{and } X^2 + Y^2 + Z^2 = v^2$$

then

$$2dX ddX + 2dY ddY + 2dZ ddZ = d(dv^2 + v^2 d\emptyset^2) \text{ and } XdX + YdY + ZdZ = vdv.$$

Hence, with the reminded calculation, the following equation will be obtained:

$$d(dv^2 + v^2 d\emptyset^2) = -2 \propto dt^2 \left[ \frac{A+C}{v^2} dv + \frac{B}{w^3} vdv - Bu \left( dX \, cos\theta + dY \, sin\theta \right) \left( \frac{1}{w^3} - \frac{1}{u^3} \right) \right]$$

where the formula $dX \, cos\theta + dY \, sin\theta$ can be conveniently expanded. Truly, considering the formulas in § 10.a above, we have that

$$dX = \frac{XdZ}{Z} - \frac{vd\emptyset \, cos\psi}{sin\sigma} \quad \text{and} \quad dY = \frac{YdZ}{Z} - \frac{vd\emptyset \, sin\psi}{sin\sigma}$$

because $Z = v \, sin\sigma \, sin\omega$, whence

$$dX \, cos\theta + dY \, sin\theta = \frac{dZ}{Z}(X \, cos\theta + Y \, sin\theta) - \frac{vd\emptyset \, cos(\theta - \psi)}{sin\sigma},$$

however, we recently saw that

$$X \, cos\theta + Y \, sin\theta = v[cos\sigma \, cos(\theta - \psi) + sin\sigma \, cos\omega \, sin(\theta - \psi)] = v \, cos\mu,$$

and since, in fact,

$$\frac{dZ}{Z} = \frac{dv}{v} + \frac{d\sigma \, cos\sigma}{sin\sigma} + \frac{d\omega \, cos\omega}{sin\omega}, \quad \text{or}$$

$$\frac{dZ}{Z} = \frac{dv}{v} + \frac{d\sigma \, cos\sigma}{sin\sigma} + \frac{d\psi \, cos\omega}{tan\sigma} \quad \text{because} \quad d\omega = \frac{d\psi \, sin\omega}{tan\sigma}.$$

---

[14] This figure was added by the Translator.

Since $d\emptyset = d\sigma + d\psi\,cos\omega$, then we get

$$\frac{dZ}{Z} = \frac{dv}{v} + \frac{d\emptyset}{tan\sigma},$$

therefore

$$dX\,cos\theta + dY\,sin\theta = dv\,cos\mu + \frac{vd\emptyset\,cos\mu}{tan\sigma} - \frac{vd\emptyset\,cos(\theta - \psi)}{sin\sigma} =$$

$$dv\,cos\mu + \frac{vd\emptyset}{sin\sigma}[sin\sigma\,cos\sigma\,cos\omega\,sin(\theta - \psi) + cos^2\sigma\,cos(\theta - \psi) - cos(\theta - \psi)]$$

and thus,

$$dX\,cos\theta + dY\,sin\theta = dv\,cos\mu - vd\emptyset[sin\sigma\,cos(\theta - \psi) - cos\sigma\,cos\omega\,sin(\theta - \psi)].$$

Therefore, the equation that we found transforms into:

$$d(dv^2 + v^2 d\emptyset^2) = -2 \propto dt^2 dv\left[\frac{A+C}{v^2} + \frac{Bv}{w^3} - Bu\,cos\mu\left(\frac{1}{w^3} - \frac{1}{u^3}\right) - \right] - 2$$

$$\propto dt^2 d\emptyset \cdot uv[sin\sigma\,cos(\theta - \psi) - cos\sigma\,cos\omega\,sin(\theta - \psi)]\left(\frac{1}{w^3} - \frac{1}{u^3}\right).$$

## XVII.

It is possible to deduce a more concise form for the expansion of the too complicated formula $dX\,cos\theta + dY\,sin\theta$ from the proper values found for $X$ and $Y$. In fact these differentials are duly produced in case where the angles $\psi$ and $\omega$ are handled as constants and $d\sigma$ is brought to $d\emptyset$, then this differential give:

$$dX = dv\,(cos\sigma\,cos\psi - sin\sigma\,cos\omega\,sin\psi) - vd\emptyset(sin\sigma\,cos\psi + cos\sigma\,cos\omega\,sin\psi)$$

$$dY = dv\,(cos\sigma\,sin\psi + sin\sigma\,cos\omega\,cos\psi) - vd\emptyset(sin\sigma\,sin\psi - cos\sigma\,cos\omega\,cos\psi)$$

whence, it follows that

$$dX\,cos\theta + dY\,sin\theta$$
$$= dv[cos\sigma\,cos(\theta - \psi) + sin\sigma\,cos\omega\,sin(\theta - \psi)]$$
$$- vd\emptyset[sin\sigma\,cos(\theta - \psi) - cos\sigma\,cos\omega\,sin(\theta - \psi)]$$

to which I observe a more contracted form in Fig. 2 where $NS = \theta - \psi$; $NZ = \sigma$, $SNZ = \omega$ and $SZ = \mu$, we will have, in the first place, as before: $cos\sigma\,cos(\theta - \psi) + sin\sigma\,cos\omega\,sin(\theta - \psi) = cos\mu$, then, in fact, if the angle $NZS = \xi$ it is found that

$$cot\,\xi = \frac{sin\sigma\,cos(\theta - \psi) - cos\sigma\,cos\omega\,sin(\theta - \psi)}{sin\omega\,sin(\theta - \psi)} \quad {}^{15}$$

whence it is concluded that

$$sin\sigma\,cos(\theta - \psi) - cos\sigma\,cos\omega\,sin(\theta - \psi) = \frac{sin\omega\,sin(\theta - \psi)\,cos\xi}{sin\xi} = sin\mu\,\,cos\xi$$

because: $sin\,\xi : sin(\theta - \psi) = sin\omega : sin\mu.$[16]

---

[15] This result comes from the application of the law of the tangent to the spherical triangle of the figure.
[16] This result comes from the application of the law of sines to the spherical triangle of the figure.

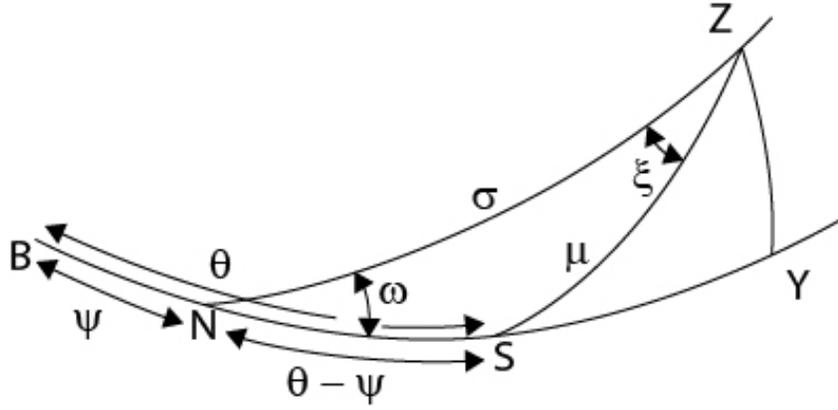

From these results we obtain

$$dX \, cos\theta + dY \, sin\theta = dv \, cos\mu - vd\emptyset \, sin\mu \, cos\xi.$$

Or if in $Z$ we draw in the direction of the arch $NZ$ another normal arch, and on it, and from $S$, we draw the perpendicular to the spherical surface, which we call $= v$, then $sinv = sin\mu \, cos\xi$,[18] or

$$sin\sigma \, cos(\theta - \psi) - cos\sigma \, cos\omega \, sin(\theta - \psi) = sinv$$

and therefore, the equation containing the first determination will assume the following form

$$d(dv^2 + v^2d\emptyset^2) = -2 \propto dt^2 \left[ \frac{A+C}{v^2}dv + \frac{Bvdv}{w^3} - Bu \, (dv \, cos\mu - vd\emptyset \, sinv) \left( \frac{1}{w^3} - \frac{1}{u^3} \right) \right].$$

XVIII.

The two remaining determinations from the differential equations deducted from the principles of motion will be conveniently obtained by the following procedures: firstly, from the equations obtained in § 14, the subtraction of the first equation multiplied by $Y$ from the second equation multiplied by $X$ gives:

$$XddY - YddX = - \propto dt^2 Bu(Y \, cos\theta - X \, sin\theta) \left( \frac{1}{w^3} - \frac{1}{u^3} \right)$$

or, when the values for $X$ and $Y$ are substituted into this expression results in

$$XddY - YddX = \propto Bvudt^2 [cos\sigma \, sin(\theta - \psi) - sin\sigma \, cos\omega \, cos(\theta - \psi)] \left( \frac{1}{w^3} - \frac{1}{u^3} \right).$$

Thus, since $XddY - YddX$ is the differential of $XdY - YdX$, then we have this equation

$$d(v^2d\emptyset \, cos\omega) = \propto Bvudt^2 [cos\sigma \, sin(\theta - \psi) - sin\sigma \, cos\omega \, cos(\theta - \psi)] \left( \frac{1}{w^3} - \frac{1}{u^3} \right).$$

In a similar way, from the first and third [equations of § 14] we deduce

$$XddZ - ZddX = - \propto BuZdt^2 cos\theta \left( \frac{1}{w^3} - \frac{1}{u^3} \right)$$

or

$$XddZ - ZddX = - \propto Bvudt^2 cos\theta \, sin\sigma \, sin\omega \left( \frac{1}{w^3} - \frac{1}{u^3} \right)$$

---

[17] This figure was added by the Translator.
[18] This geometrical construction is not clear.

then giving

$$d(v^2 d\emptyset \sin\omega \cos\psi) = -\propto Bvudt^2 \cos\theta \sin\sigma \sin\omega \left(\frac{1}{w^3} - \frac{1}{u^3}\right).$$

Equally, from the second equation combined with the third equation results in

$$YddZ - ZddY = -\propto Bvudt^2 \sin\theta \sin\sigma \sin\omega \left(\frac{1}{w^3} - \frac{1}{u^3}\right)$$

or

$$d(v^2 d\emptyset \sin\omega \sin\psi) = -\propto Bvudt^2 \sin\theta \sin\sigma \sin\omega \left(\frac{1}{w^3} - \frac{1}{u^3}\right)$$

where it should be noted that only two determinations are contained in these three equations, and the third [equation] can be freely included with the two other [equations].

<div align="center">XIX.</div>

So, let us examine the two last [equations], and since their first members ought to be differentiate, and observing that the quantity $v^2 d\emptyset \sin\omega$ is a unique quantity to both, then

$$\cos\psi \, d(\,v^2 d\emptyset \sin\omega) - d\psi \sin\psi \, v^2 d\emptyset \sin\omega = -\propto Bvudt^2 \cos\theta \sin\sigma \sin\omega \left(\frac{1}{w^3} - \frac{1}{u^3}\right)$$

$$\sin\psi \, d(\,v^2 d\emptyset \sin\omega) + d\psi \cos\psi \, v^2 d\emptyset \sin\omega = -\propto Bvudt^2 \sin\theta \sin\sigma \sin\omega \left(\frac{1}{w^3} - \frac{1}{u^3}\right)$$

whence, eliminating $d(\,v^2 d\emptyset \sin\omega)$ results in

$$d\psi \, v^2 d\emptyset \sin\omega = -\propto Bvudt^2 \sin\sigma \sin\omega \sin(\theta - \psi) \left(\frac{1}{w^3} - \frac{1}{u^3}\right)$$

and thus the variation of the nodal line is defined such that

$$d\psi = \frac{-\propto Budt^2 \sin\sigma \, \sin(\theta - \psi)}{v \, d\emptyset} \left(\frac{1}{w^3} - \frac{1}{u^3}\right)$$

from which, at the same time, the variation of the inclination is obtained from $\frac{d\omega}{\sin\omega} = \frac{d\psi}{\tan\sigma}$, and, on the other hand, once the member $v^2 d\emptyset \sin\omega$ is eliminated from both original equations, the following equation is obtained

$$d(\,v^2 d\emptyset \sin\omega) = -\propto Bvudt^2 \sin\sigma \sin\omega \cos(\theta - \psi) \left(\frac{1}{w^3} - \frac{1}{u^3}\right).$$

Writing the first equation in the following form:

$$\cos\omega \, d(v^2 d\emptyset) - d\omega \sin\omega \, v^2 d\emptyset = \propto Bvudt^2 [\cos\sigma \sin(\theta - \psi) - \sin\sigma \cos\omega \cos(\theta - \psi)] \left(\frac{1}{w^3} - \frac{1}{u^3}\right),$$

and expanding the first member of the last equation, gives

$$\sin\omega \, d(\,v^2 d\emptyset) + d\omega \cos\omega \, v^2 d\emptyset = -\propto Bvudt^2 [\sin\sigma \sin\omega \cos(\theta - \psi)] \left(\frac{1}{w^3} - \frac{1}{u^3}\right).$$

Eliminating $d\omega$ between the last two equations, gives

$$d(\,v^2 d\emptyset) = -\propto Bvudt^2 dt^2 [\sin\sigma \cos(\theta - \psi) - \cos\sigma \cos\omega \sin(\theta - \psi)] \left(\frac{1}{w^3} - \frac{1}{u^3}\right)$$

or

$$d(\,v^2 d\phi\,) = -\propto Bvudt^2 \, sinv\left(\frac{1}{w^3} - \frac{1}{u^3}\right)$$

which is the other determination required to be sought.

XX.

Let us multiply this last equation by $2v^2 d\phi$, and leaving one integral just indicated, we will have the following expression

$$v^4 d\phi^2 \; = -\mathbf{2} \propto Bdt^2 \int v^3 u d\phi \, u \, sinv\left(\frac{1}{w^3} - \frac{1}{u^3}\right) \quad [19]$$

this equation contains the relation between the elementary angle $d\phi$ and the infinitesimal time $dt$, where, in fact, it is clear that if the mass of de disturbing body $B$ would fade away, then $v^2 d\phi$ would be proportional to the time $dt$, or the areas described around $A$ are proportional to time. If to this equation, it is first added the one found in § 17, and equally integrated to the extent possible, then

$$dv^2 + v^2 d\phi^2 = 2 \propto dt^2 (A+C)\left(\frac{1}{v} - \frac{1}{f}\right) - 2$$
$$\propto Bdt^2 \int \frac{v dv}{w^3} + 2 \propto Bdt^2 \int u\,(dv\,cos\mu - vd\phi\,sinv)\left(\frac{1}{w^3} - \frac{1}{u^3}\right).$$

the above equation compares the variation of the distance $v$ with the element $d\phi$ or with the infinitesimal time $dt$, which are two particular characteristics to look for the motion of the body $Z$ in its own orbit. Besides, we have, in fact, for the variation of orbit itself the following:

$$d\psi = \frac{-\propto Budt^2 sin\sigma \; sin(\theta - \psi)}{v\,d\phi}\left(\frac{1}{w^3} - \frac{1}{u^3}\right)$$

$$\frac{d\omega}{sin\omega} = \frac{-\propto Budt^2 cos\sigma \; sin(\theta - \psi)}{v\,d\phi}\left(\frac{1}{w^3} - \frac{1}{u^3}\right) = \frac{d\psi}{tan\sigma}.$$

And finally, it can be recalled the relation between the argument of the latitude $\sigma$ to these same elements, given by $d\sigma = d\phi - d\psi\,cos\omega.$

XXI.

The element of time $dt$ with the constant $\propto$ will be taken away from the calculation in a most convenient way, if a certain motion that is regular and known is introduced, like the mean motion of the Sun, or of another body, which is revolved around the center of the forces in a uniform circle. Then, let us put around the body located in $A$, which mass is $= \mathfrak{A}$, the other body, which mass $= \mathfrak{C}$, at a distance $= a$, in a circle, so that it circulates in a time $t$ an angle $\tau$ proportional to it, and our equations can be adapted to this case, once these are established: $A = \mathfrak{A}, C = \mathfrak{C}$ and $B = 0$, and then $v = a$ and $d\phi = d\tau$. Then, the motion that we assume to be known, is controlled by these two equations

$$v^4 d\phi^2 \; = \mathbf{2} \propto Ddt^2 \qquad \text{and} \qquad dv^2 + v^2 d\phi^2 = 2 \propto dt^2(\mathfrak{A} + \mathfrak{C})\left(\frac{1}{v} - \frac{1}{f}\right)$$

---

[19] The number two in bold should not be there.

where the first constants $D$ and $f$ can be conveniently defined for this case. To this end, from the first $\mathbf{2} \propto dt^2 = \frac{v^4 d\emptyset^2}{D}$ which, when substituted into the second, gives $dv^2 + v^2 d\emptyset^2 = \frac{(\mathfrak{A}+\mathfrak{C})v^4 d\emptyset^2}{D}\left(\frac{1}{v}-\frac{1}{f}\right)$ or $Df dv^2 + Df v^2 d\emptyset^2 = (\mathfrak{A} + \mathfrak{C})v^3 d\emptyset^2(f - v)$, from which

$$d\emptyset = \frac{dv\sqrt{Df}}{v\sqrt{(\mathfrak{A}+\mathfrak{C})v(f-v) - Df}},$$

a certain constant value of $v$ itself should be satisfactorily attributed to this differential equation, for which the denominator fades away, however, as I have exposed in another place, this approach is not possible to be admitted to the integral, unless the factor of denomination[20] fades away to be of a minimum dimension of one, whence it is necessary that under the radical sign the same factor appears in a pair or squared, such that it reduces to the same value, so that the differential of the quantity placed after the sign reproduces the same factor. Therefore, let us place that differential $= 0$, and $v = \frac{1}{2}f$, which under the hypothesis that $v = a$, then $f = 2a$, and in this case the denominator itself is then equal to $(\mathfrak{A} + \mathfrak{C})a^2 - 2Da$, which, when equated to zero, gives $D = \frac{1}{2}(\mathfrak{A} + \mathfrak{C})a$. Now, in the other equation be considered that $v = a$ and $d\emptyset = d\tau$, then $a^4 d\tau^2 = \propto (\mathfrak{A} + \mathfrak{C})a dt^2$ or $\propto dt^2 = \frac{a^3 d\tau^2}{\mathfrak{A}+\mathfrak{C}}$.

<center>XXII.</center>

Then, since the knowledge of the motion at any given time $t$ allows the determination of the mean motion $\tau$, here the time variable in our calculation will be redefined, by writing in the place of $\propto dt^2$ the value just found of $\frac{a^3 d\tau^2}{\mathfrak{A}+\mathfrak{C}}$. Next, to render our formulas simpler, let us put $\frac{A+C}{\mathfrak{A}+\mathfrak{C}} = m$ and $B = n(A + C)$, and then $\propto dt^2(A + C) = ma^3 d\tau^2$ and $\propto B dt^2 = mna^3 d\tau^2$ where it should be noted that the perturbations will be minimum if the terms affected by the number $n$ are minimum. Then our equations will assume the following forms:

1º. $v^4 d\emptyset^2 = -\mathbf{2}mna^3 d\tau^2 \int v^3 u d\emptyset \ sinv\left(\frac{1}{w^3}-\frac{1}{u^3}\right)$

2º. $dv^2 + v^2 d\emptyset^2 =$

$$2ma^3 d\tau^2\left(\frac{1}{v}-\frac{1}{f}\right) - 2mna^3 d\tau^2 \int \frac{v dv}{w^3} + 2mna^3 d\tau^2 \int u \ (dv \ cos\mu - v d\emptyset \ sinv)\left(\frac{1}{w^3}-\frac{1}{u^3}\right)$$

3º. $d\psi = -mna^3 d\tau^2 \frac{u \ sin\sigma \ sin(\theta-\psi)}{v \ d\emptyset}\left(\frac{1}{w^3}-\frac{1}{u^3}\right)$

4º. $\frac{d\omega}{sin\omega} = -mna^3 d\tau^2 \frac{u \ cos\sigma \ sin(\theta-\psi)}{v \ d\emptyset}\left(\frac{1}{w^3}-\frac{1}{u^3}\right) = \frac{d\psi}{tan\sigma}$

5º. $d\sigma = d\emptyset + mna^3 d\tau^2 cos\omega \frac{u sin\sigma \ sin(\theta-\psi)}{v \ d\emptyset}\left(\frac{1}{w^3}-\frac{1}{u^3}\right) = d\emptyset - d\psi \ cos\omega$ .

With these equations, the whole motion of the body $Z$ with all the perturbations originated by the action of the body $S$ can be determined: where, especially regarded to the integral formulas, which in the first two equations are affected only by the perturbations, being sufficient that these values are selected as close to the real ones, from which the task of approximations to these integrals can hardly be considered an impediment. Nonetheless, I will expose soon the method to such an extent as to liberate the calculation of these integrals.

---

[20] It is simply the result of the division of the ratio.

Meanwhile, for the sake of brevity, let us consider that:

$$\int v^3 u d\emptyset \, u \, sinv \left(\frac{1}{w^3} - \frac{1}{u^3}\right) = P$$

$$\int \frac{vdv}{w^3} = Q$$

$$\int u \, (dv \, cos\mu - vd\emptyset \, sinv) \left(\frac{1}{w^3} - \frac{1}{u^3}\right) = R$$

then, the first two previous equations are contracted to these forms:

$$1^{\underline{o}}. \; v^4 d\emptyset^2 = 2ma^3 d\tau^2 (D - nP)$$

$$2^{\underline{o}}. \; dv^2 + v^2 d\emptyset^2 = 2ma^3 d\tau^2 \left[\frac{1}{v} - \frac{1}{f} - n(Q - R)\right]$$

which alone accomplishes all the task, if the body $Z$ is moved in the same plane, in which we assume that the disturbing body $S$ is circulating, the remaining equations for the motion, that is declared to pertain to the latitude, the solution of these suffers much less difficulties, since all the efforts should be consumed in the two previous [equations]. Thenceforth, once the element $d\tau$ has been eliminated, arises this equation

$$(D - nP)(dv^2 + v^2 d\emptyset^2) = v^4 d\emptyset^2 \left[\left(\frac{1}{v} - \frac{1}{f}\right) - n(Q - R)\right]$$

whence the following equation is obtained

$$d\emptyset = \frac{dv\sqrt{(D - nP)}}{v\sqrt{\left(v - \frac{v^2}{f} - nv^2(Q - R) - D + nP\right)}}$$

and furthermore

$$2ma^3 d\tau^2 = \frac{v^2 dv^2}{v - \frac{v^2}{f} - nv^2(Q - R) - D + nP}$$

or

$$ad\tau\sqrt{2ma} = \frac{vdv}{\sqrt{-D + nP + v - v^2\left(\frac{1}{f} + nQ - nR\right)}}$$

the integration of these formulas would be available in case the fraction $n$ or the perturbations disappeared.

We can represent that equation in this form:

$$\frac{dv}{v^2}\sqrt{(D - nP)} = d\emptyset\sqrt{\left(-\frac{1}{f} - n(Q - R) + \frac{1}{v} - \frac{D - nP}{v^2}\right)}$$

and since we know that $AZ = v$, then, it would become maximum or minimum when the quantity under the radical sign vanishes. However, not only in Astronomy that these places are of primary importance, wherever the body $Z$ is said to move along a segment of an arch, but even so, the choice of this important fact is at our disposal, with which the disturbed motion could be very neatly compared with the regular motion, and thus be capable to assign the aberrations from it. However, in a convenient way, this will be provided by introducing into the calculation a new angle $\Upsilon$, which in astronomy is called the true anomaly[21], and it is chosen in such way that either by reducing or increasing the [angular] distance between two lines according to the maximum or minimum value of $v$. Therefore, with the purpose of approximating the real motion to a regular motion made along an ellipse, we now define $v = \frac{p}{1 + q\,cos\Upsilon}$, so that now the motion conforms to the regular motion along such an ellipse, in which the semi-latus rectum is $= p$, the eccentricity $= q$, and thus the semi-major axis $= \frac{p}{1-q^2}$, and the true anomaly arising from the [major] axis $= \Upsilon$. Then, it can be easily perceived that because of the perturbations, the aspect of this ellipse changes continuously, whence not only the anomaly $\Upsilon$ but also the letters $p$ and $q$ are expected to vary, and these variations are now investigated.

<p style="text-align:center">XXV.</p>

For that investigation to be rendered easier, let us introduce, for the sake of brevity, the following:

$$\frac{1}{f} - n(Q - R) = M \quad \text{and} \quad D - nP = N$$

such that we have this evolved form

$$\frac{dv}{v^2}\sqrt{(D - nP)} = d\emptyset\sqrt{\left(-M + \frac{1}{v} - \frac{N}{v^2}\right)}.$$

Now that we have $v = \frac{p}{1 + q\,cos\Upsilon}$ or $\frac{1}{v} = \frac{1 + q\,cos\Upsilon}{p}$, by hypothesis, not only for the case where $\Upsilon = 0$, which gives $\frac{1}{v} = \frac{1+q}{p}$, as well as for the case where $\Upsilon = 180^\circ$, which gives $\frac{1}{v} = \frac{1-q}{p}$, the quantity $-M + \frac{1}{v} - \frac{N}{v^2}$ should go off to zero[22], and then, from these two situations arise the equations:

$$-M + \frac{1+q}{p} - \frac{N(1+q)^2}{p^2} = 0$$

and

$$-M + \frac{1-q}{p} - \frac{N(1-q)^2}{p^2} = 0$$

and by subtraction they give $\frac{2q}{p} - \frac{4Nq}{p^2} = 0$, such that $p = 2N$, or $N = \frac{1}{2}p$, whence $M = \frac{1+q}{p} - \frac{(1+q)^2}{2p} = \frac{1-q^2}{2p}$.

But if we put the semi-major axis of our ellipse $= r$ such that $r = \frac{p}{1-q^2}$, then $M = \frac{1}{2r}$, and thus

$$\frac{1}{f} - n(Q - R) = \frac{1}{2r} \quad \text{and} \quad D - nP = \frac{1}{2}p\,.$$

---

[21] The true anomaly (Υ) represents the real geometric angle in the plane of the ellipse, between periapsis (closest approach to the central body) and the position of the orbiting object at any given time. Argument of periapsis (β) defines the orientation of the ellipse in the orbital plane, as an angle measured from the ascending node to the periapsis (the closest point the celestial body [e.g Moon] comes to the central body [e.g Earth] around which it orbits).

[22] By equating this quantity to zero, Euler is somehow searching for the maximum and minimum values of $v$, which translates into finding the major and minor axes of the ellipse.

When these values are substituted in our equation results in:

$$\frac{dv}{v^2}\sqrt{\frac{1}{2}p} = d\emptyset\sqrt{\left(\frac{-1+q^2}{2p} + \frac{1}{v} - \frac{p}{2v^2}\right)}$$

and when we write for $\frac{1}{v}$ in the right-hand side of this equation, the value $\frac{1+q\,cos\Upsilon}{p}$, then we have that:

$$\frac{dv}{v^2}\sqrt{\frac{1}{2}p} = d\emptyset\sqrt{\left(\frac{-1+q^2}{2p} + \frac{1+q\,cos\Upsilon}{p} - \frac{1+2q\,cos\Upsilon+q^2cos^2\Upsilon}{2p}\right)} \quad \text{23}$$

and thus,

$$\frac{dv}{v^2}\sqrt{\frac{1}{2}p} = d\emptyset\sqrt{\frac{q^2-q^2cos^2\Upsilon}{2p}} = \frac{qd\emptyset\,sin\Upsilon}{\sqrt{2p}}$$

So, we have that $\frac{dv}{v^2} = \frac{qd\emptyset}{p}sin\Upsilon$; whence, surely, as we have anticipated, we certainly recognize that as the anomaly fades away with the sine of $\Upsilon$, at the same time, the differential of $v$ goes off to zero, and, therefore, it passes over a maximum or a minimum value. Hence, indeed, the increment of the distance $v$, in general, is reduced to the element $d\emptyset$, which itself can now be compared with the known element $d\tau$, and because $D - nP = \frac{1}{2}p$, we have that

$$v^4d\emptyset^2 = ma^3pd\tau^2, \qquad \text{or} \qquad v^2d\emptyset = ad\tau\sqrt{map}.$$

And since $\frac{1}{v} = \frac{1+q\,cos\Upsilon}{p}$, then

$$\frac{dv}{v^2} = \frac{dp(1+q\,cos\Upsilon)}{p^2} - \frac{dq\,cos\Upsilon + q\,d\Upsilon\,sin\Upsilon}{p}$$

which turned out equal to the expression $\frac{qd\emptyset}{p}sin\Upsilon$, resulting in

$$q(d\emptyset - d\Upsilon)sin\Upsilon = \frac{dp}{p}(1+q\,cos\Upsilon) - dq\,cos\Upsilon = \frac{dp}{v} - dq\,cos\Upsilon,$$

which involves new relations among differentials.

The remaining determinations must be sought from the formulas found above:

$$p = 2D - 2nP \quad \text{and} \quad \frac{1}{r} = \frac{1-q^2}{p} = \frac{2}{f} - 2n(R-Q),$$

which once differentiated, and substituting the restituted values shown above for $P, Q, R$, give

$$dp = -2ndP = -2nv^3ud\emptyset\,sinv\left(\frac{1}{w^3} - \frac{1}{u^3}\right)$$

$$d\left(\frac{1}{r}\right) = d\left(\frac{1-q^2}{p}\right) = 2ndQ - 2ndR = \frac{2nvdv}{w^3} - 2nu\,(dv\,cos\mu - vd\emptyset\,sinv)\left(\frac{1}{w^3} - \frac{1}{u^3}\right).$$

---

[23] There are errors in the signs of the last term under the radical which were corrected.

However, since $dv = \frac{qv^2 d\emptyset sin\Upsilon}{p} = \frac{qvd\emptyset \, sin\Upsilon}{1+q\,cos\Upsilon}$, then, the last differential reduced to the element $d\emptyset$ transforms into

$$d\left(\frac{1}{r}\right) = d\left(\frac{1-q^2}{p}\right) = \frac{2nqv^3 d\emptyset \, sin\Upsilon}{pw^3} - 2nvud\emptyset \left(\frac{q\,cos\mu\,sin\Upsilon}{1+q\,cos\Upsilon} - sinv\right)\left(\frac{1}{w^3} - \frac{1}{u^3}\right).$$

It is true that $d\left(\frac{1-q^2}{p}\right) = \frac{-dp}{p^2}(1-q^2) - \frac{2qdq}{p}$, and then

$$qdq = -\frac{dp}{2p}(1-q^2) - \frac{p}{2}d\left(\frac{1-q^2}{p}\right),$$

from which it is conclude that

$$qdq = +\frac{n(1-q^2)}{p}v^3ud\emptyset \, sinv\left(\frac{1}{w^3} - \frac{1}{u^3}\right) - \frac{nqv^3 d\emptyset \, sin\Upsilon}{w^3}$$

$$+ npvud\emptyset \left(\frac{q\,cos\mu\,sin\Upsilon}{1+q\,cos\Upsilon} - sinv\right)\left(\frac{1}{w^3} - \frac{1}{u^3}\right)$$

which, since $v = \frac{p}{1+q\,cos\Upsilon}$, it is contracted into the following form

$$qdq = nv^2ud\emptyset\left(\frac{1}{w^3} - \frac{1}{u^3}\right)\left[q\,cos\mu\,sin\Upsilon - \frac{q\,sinv\,(q+2\,cos\Upsilon + q\,cos^2\Upsilon)}{1+q\,cos\Upsilon}\right] - \frac{nqv^3 d\emptyset \, sin\Upsilon}{w^3},$$

which, once divided by $q$, gives

$$dq = nv^2ud\emptyset\left(\frac{1}{w^3} - \frac{1}{u^3}\right)\left[cos\mu\,sin\Upsilon - \frac{sinv\,(q+2\,cos\Upsilon + q\,cos^2\Upsilon)}{1+q\,cos\Upsilon}\right] - \frac{nv^3 d\emptyset \, sin\Upsilon}{w^3}.$$

Finally, when this expression is substituted into the formula $q(d\emptyset - d\Upsilon)sin\Upsilon = \frac{dp}{v} - dq\,cos\Upsilon$, the resulting expression, once divided by $sin\Upsilon$, is

$$q(d\emptyset - d\Upsilon) = \frac{nv^3 d\emptyset \, cos\Upsilon}{w^3} - nv^2ud\emptyset\left(\frac{1}{w^3} - \frac{1}{u^3}\right)\left[cos\mu\,cos\Upsilon + \frac{sinv\,sin\Upsilon(2+qcos\Upsilon)}{1+q\,cos\Upsilon}\right].$$

XXVII.

Therefore, now we have all the quantities that enters into our calculation, and we have revealed the sudden increment in the element $d\emptyset$, which in the same infinitesimal time $dt$ is completed by the here introduced angle $d\tau$ according to the mean motion , whence, we can easily assign that increment for any minimum time. Hence, firstly, the relation between the elementary angle $d\emptyset$ and $d\tau$ is expressed by the following formula:

$$v^2 d\emptyset = ad\tau\sqrt{map} \quad \text{whence, we have that} \quad ma^3 d\tau^2 = \frac{1}{p}v^4 d\emptyset^2$$

next, if we consider that $v = \frac{p}{1+q\,cos\Upsilon}$, and that $r = \frac{p}{1-q^2}$, then we will have that:

$$1°.\ dp = -2nv^3ud\emptyset \, sinv\left(\frac{1}{w^3} - \frac{1}{u^3}\right)$$

$$2°.\ d\left(\frac{1}{r}\right) = \frac{2nqv^3 d\emptyset \, sin\Upsilon}{pw^3} - \frac{2nv^2ud\emptyset}{p}[q\,cos\mu\,sin\Upsilon - (1+q\,cos\Upsilon)sinv]\left(\frac{1}{w^3} - \frac{1}{u^3}\right)$$

$$3^{\circ}.\ dq = nv^2 ud\phi \left(\frac{1}{w^3} - \frac{1}{u^3}\right)\left[cos\mu\ sin\Upsilon - \frac{(q + 2\ cos\Upsilon + q\ cos^2\Upsilon)\ sinv}{1 + q\ cos\Upsilon}\right] - \frac{nv^3 d\phi\ sin\Upsilon}{w^3}$$

$$4^{\circ}.\ d\Upsilon = d\phi - \frac{nv^3 d\phi\ cos\Upsilon}{qw^3} - \frac{nv^2 ud\phi}{q}\left(\frac{1}{w^3} - \frac{1}{u^3}\right)\left[cos\mu\ cos\Upsilon + \frac{(2 + qcos\Upsilon)sinv\ sin\Upsilon}{1 + q\ cos\Upsilon}\right],$$

knowing that

$$cos\mu = cos\sigma\ cos(\theta - \psi) + sin\sigma\ cos\omega\ sin(\theta - \psi)$$

and that

$$sinv = sin\sigma\ cos(\theta - \psi) - cos\sigma\ cos\omega\ sin(\theta - \psi)$$

where it should be noted that $d\phi - d\Upsilon$ designates the increment of the arc described by the celestial body along the orbit itself.

Third, for the motion in the latitude we have these formulas:

$$1^{\circ}.\ d\psi = -\frac{nv^3 ud\phi\ sin\sigma\ sin(\theta - \psi)}{p}\left(\frac{1}{w^3} - \frac{1}{u^3}\right)$$

$$2^{\circ}.\ \frac{d\omega}{sin\omega} = -\frac{nv^3 ud\phi\ cos\sigma\ sin(\theta - \psi)}{p}\left(\frac{1}{w^3} - \frac{1}{u^3}\right) = \frac{d\psi}{tan\sigma}$$

$$3^{\circ}.\ d\sigma = d\phi + \frac{nv^3 ud\phi\ cos\omega\ sin\sigma\ sin(\theta - \psi)}{p}\left(\frac{1}{w^3} - \frac{1}{u^3}\right)$$

or $d\sigma = d\phi - d\psi\ cos\omega$ .

XXIX.

Nothing more would be desired, if I could perform the integration of these equations, since then, any kind of perturbation could be defined, no matter great it would have been. But, since the forces of the Analysis have not yet increased to such an extent, it is fit to appeal to approximations, from which, hopefully, we could expect some success, in case the perturbations would be considered as being small: because then, the values of the quantities $p$ and $q$ would be changed by small amounts, having been affected by the letter $n$ in the integration of the formulas, such that, without error, can be considered just as constants, since, indeed, later on, the necessary corrections, obtained by the usual methods, can be applied without difficulties. However, in case we consider that the eccentricity is rather big, we will have difficulties, which, however, in order to be overcome, it will be possible to apply certain artifices, in which, indeed, the best way to succeed, is to consider that the eccentricity $q$, varies little from unity, such as in an almost parabolic orbit as those of the comets. Nonetheless, greater difficulties appear, when the eccentricity is rather small, especially when the variations of the anomaly $\Upsilon$ grow, however if the need should arise, this cure could be expected. These operations are chiefly hindered by the quantity $\frac{1}{w^3}$, unless it is possible to conveniently convert it into a series sufficiently convergent, whose whole integration would be despairing, and another option is not seen to be left over, unless from the single variations of these differential formulas versus very small intervals of time are defined, the task for the integration of these summations is compensated, and on another occasion I showed more details.

APPLICATION OF THIS THEORY TO THE MOTION OF THE MOON

XXX.

Let us consider that the center of the Earth is in $A$, which mass is $= A$, and considering the plane of the ecliptic in the figure, now, indeed, we have the Sun in $S$, which mass is $= B$, and for which, I place the line $AB$ directed to a fixed point in the sky such as to the First Star of Aries, being defined the following

<div align="center">longitude of the Sun or angle $BAN = \theta$</div>

<div align="center">and distance of the Sun to the Earth or $AS = u$.</div>

These elements are defined for the Sun's motion: semi-major axis of the Sun's orbit $= a$, semi-latus rectum of the Sun's orbit $= b$, eccentricity of its orbit $= e$, and true anomaly $= v$; then $u = \frac{b}{1-e\,cosv}$.[24] Then, in fact, considering the mean motion[25], if during an infinitesimal time the Sun traverses an angle $d\tau$, then $u^2 d\theta = a d\tau \sqrt{ab}$,[26] where $d\theta = dv$, and we wish to observe the motion only during the Sun's apogee. It is true that $b = a(1 - e^2)$; then we have that

$$u = \frac{a(1-e^2)}{1-e\,cosv}, \quad \text{and} \quad d\tau = \frac{d\theta(1-e^2)^{\frac{3}{2}}}{(1-e\,cosv)^2} \quad \text{and also,}\, dv = d\theta, \text{approximately,}$$

where it should be noted that once the eccentricity of the Sun's orbit is neglected, then $u = a$ and $d\theta = d\tau$.

<div align="center">XXXI.</div>

Be further the Moon in $Z$, which mass $= C$, and be defined $\frac{B}{A+C} = n$, knowing that $\frac{A+C}{A+B} = m$, and since the elementary angle $d\tau$ is chosen from the mean motion of the Sun, such that $\mathfrak{A} = A$ and $\mathfrak{C} = B$; whence giving $\frac{B}{A+B} = mn$ or $m = \frac{1}{n'}$, since the mass of the Sun can be considered infinitely larger than the mass of the Earth. Now for the location of the Moon it is established that

<div align="center">longitude of the ascending node or angle $BAN = \psi$</div>

<div align="center">inclination of the orbit of the Moon in relation to the ecliptic or angle $YOZ = \omega$</div>

<div align="center">and argument of latitude or angle $NAZ = \sigma$</div>

then, we have that:

<div align="center">longitude of the Moon$= \psi + \underbrace{arctan(tan\sigma\ cos\omega)}_{this\ is\ the\ angle\ NAY}$</div>

and

<div align="center">northern latitude $= \underbrace{arcsin\ (sin\sigma\ sin\omega)}_{this\ is\ the\ angle\ ZAY}.$</div>

Then, if the distance of the Moon to the Earth $AZ = v$ and the distance of the Moon to the Sun $SZ = w$, and having been already defined the two angles $\mu$ and $\nu$, such that

$$cos\mu = cos\sigma\ cos(\theta - \psi) + sin\sigma\ cos\omega\ sin(\theta - \psi)$$

$$sin\nu = sin\sigma\ cos(\theta - \psi) - cos\sigma\ cos\omega\ sin(\theta - \psi),$$

then, $w^2 = v^2 + u^2 - 2vu\ cos\mu$ or $w = \sqrt{v^2 + u^2 - 2vu\ cos\mu}$.

---

[24] The negative sign in the denominator indicates that the reference direction $\theta = 0$ points towards the center of the ellipse, and positive if that direction points away from the center.

[25] In orbital mechanics, mean motion is the angular speed required for a body to complete one orbit, assuming constant speed in a circular orbit which completes in the same time as the variable speed elliptical orbit of the actual body.

[26] This result comes from Kepler's 2nd law.

From these considerations, if we consider that currently $v = \frac{p}{1-q\,cos\Upsilon}$ denotes half the parameter of the lunar orbit, $q$ is its eccentricity, and the angle $\Upsilon$ is its true anomaly; such that now $q$ should be considered negative; then, in fact, being $d\emptyset$ the angle described by the Moon around the Earth in the same time, during which the Sun traverses the angle $d\tau$ in its mean motion. Hence, for defining the motion of the Moon, the following equations should be considered:

$$1^{\underline{o}}.\ v^2 d\emptyset = ad\tau\sqrt{map} = ad\tau\sqrt{\frac{ap}{n}}$$

$$2^{\underline{o}}.\ dp = -2nv^3 u d\emptyset\ sin\nu\left(\frac{1}{w^3}-\frac{1}{u^3}\right)$$

$$3^{\underline{o}}.\ dq = -nv^2 u d\emptyset\left[cos\mu\,sin\Upsilon+\frac{(q-2\,cos\Upsilon+q\,cos^2\Upsilon)\,sin\nu}{1-q\,cos\Upsilon}\right]\left(\frac{1}{w^3}-\frac{1}{u^3}\right)+\frac{nv^3 d\emptyset\,sin\Upsilon}{w^3}$$

$$4^{\underline{o}}.\ \frac{dv}{v^2}=-\frac{qd\emptyset\,sin\Upsilon}{p}\quad\text{or}\quad d\left(\frac{1}{v}\right)=+\frac{qd\emptyset\,sin\Upsilon}{p}$$

$$5^{\underline{o}}.\ d\Upsilon = d\emptyset+\frac{nv^3 d\emptyset\,cos\Upsilon}{qw^3}-\frac{nv^2 u d\emptyset}{q}\left[cos\mu\,cos\Upsilon+\frac{(2-q\,cos\Upsilon)sin\nu\,sin\Upsilon}{1-q\,cos\Upsilon}\right]\left(\frac{1}{w^3}-\frac{1}{u^3}\right)$$

denoting $d\emptyset - d\Upsilon$ the instantaneous promotion of the line of apsides[27] or the [promotion] of the Moon's apogee in its orbit:

$$6^{\underline{o}}.\ d\psi = \frac{-nv^3 u d\emptyset\,sin\sigma\,sin(\theta-\psi)}{p}\left(\frac{1}{w^3}-\frac{1}{u^3}\right)$$

$$7^{\underline{o}}.\ \frac{d\omega}{sin\omega}==\frac{d\psi}{tan\sigma}=\frac{-nv^3 u d\emptyset\,cos\sigma\,sin(\theta-\psi)}{p}\left(\frac{1}{w^3}-\frac{1}{u^3}\right)$$

$$8^{\underline{o}}.\ d\sigma = d\emptyset-d\psi\,cos\omega = d\emptyset+\frac{nv^3 u d\emptyset\,sin\sigma\,cos\omega\,sin(\theta-\psi)}{p}\left(\frac{1}{w^3}-\frac{1}{u^3}\right).$$

Let us begin with the development of the expression $\frac{1}{w^3}-\frac{1}{u^3}$; since it is obvious to be established that the distance $u$ is always much longer than the distance $v$; then we have that

$$\frac{1}{w^3}=(u^2-2vu\,cos\mu+v^2)^{-1/2}=\frac{1}{u^3}+\frac{3v\,cos\mu}{u^4}+\frac{3v^2\,(5\,cos^2\mu-1)}{2u^5}$$

And, therefore,

---

27

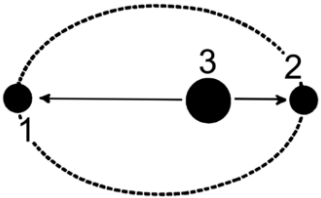

The apsides refer to the farthest (1) and nearest (2) points reached by an orbiting planetary body (1 and 2) with respect to a primary, or host, body (3). The line of apsides is the line connecting positions 1 and 2. In the case of Moon and Earth, point 1, the farthest, is called apogee; and point 2, the nearest, is called perigee.

$$\frac{1}{w^3} - \frac{1}{u^3} = \frac{3v\,cos\mu}{u^4} + \frac{3v^2\,(5\,cos^2\mu - 1)}{2u^5}.$$

Then, the formulas nº. 3 and nº. 5 will transform into these expressions:

$$3º.\ \ dq = \frac{nv^3 d\emptyset\,sin\Upsilon}{u^3}\left[1 - 3\,cos^2\mu + \frac{3v\,cos\mu}{2u}(3 - 5\,cos^2\mu)\right]$$

$$-\frac{nv^2 d\emptyset\,sinv}{1 - q\,cos\Upsilon}(q - 2\,cos\Upsilon + q\,cos^2\Upsilon)\left(\frac{3v\,cos\mu}{u^3} + \frac{3v^2\,(5\,cos^2\mu - 1)}{2u^4}\right)$$

$$5º.\ \ d\Upsilon = d\emptyset + \frac{nv^3 d\emptyset\,cos\Upsilon}{qu^3}\left[1 - 3\,cos^2\mu + \frac{3v\,cos\mu}{2u}(3 - 5\,cos^2\mu)\right]$$

$$-\frac{nv^2 ud\emptyset\,sinv\,sin\Upsilon}{q(1 - q\,cos\Upsilon)}\left(\frac{3v\,cos\mu}{u^3} + \frac{3v^2\,(5\,cos^2\mu - 1)}{2u^4}\right)(2 - qcos\Upsilon).$$

Then, we have that:

$$dq\,cos\Upsilon + q(d\emptyset - d\Upsilon)\,sin\Upsilon = 2nv^2 nd\emptyset\,sinv\left(\frac{3v\,cos\mu}{u^3} + \frac{3v^2\,(5\,cos^2\mu - 1)}{2u^4}\right)$$

which is a quite simple expression that will be possible to use next.

<div align="center">XXXIV.</div>

Then, it should be furthermore noted that when the inclination $\omega$ is quite small, such that $cos\omega = 1 - \frac{1}{2}\omega^2$, then, approximately, we have that:

$$cos\mu = \ cos(\sigma - \theta + \psi) - \frac{1}{2}\omega^2 sin\sigma\,sin(\theta - \psi)$$

$$sinv = sin(\sigma - \theta + \psi) + \frac{1}{2}\omega^2 cos\sigma\,cos\omega\,sin(\theta - \psi).$$

<div align="center">___________________________</div>